\documentstyle[12pt,prc,preprint,tighten,aps,psfig]{revtex}

\begin {document}
\draft

\title{
Unified treatment of the Coulomb and 
harmonic oscillator potentials in $D$ dimensions
}

\author{G. L\'evai, B. K\'onya and Z. Papp}

\address{
Institute of Nuclear Research of the Hungarian
Academy of Sciences, \\
P. O. Box 51, H--4001 Debrecen, Hungary 
}

\maketitle

\begin{abstract}
Quantum mechanical models and practical calculations often rely 
on some exactly solvable models like the Coulomb and the 
harmonic oscillator potentials.
The $D$ dimensional generalized Coulomb potential contains
these potentials as limiting cases, thus it establishes 
a continuous link 
between the Coulomb and harmonic oscillator potentials in 
various dimensions.
We present results which are necessary
for the utilization of this potential as a model and practical
reference problem for quantum mechanical calculations.
We define a Hilbert space basis, the generalized Coulomb-Sturmian 
basis, and calculate the Green's operator on this basis and 
also present an SU(1,1) algebra associated with it. 
We formulate the problem for the one-dimensional 
case too, and point out that the complications arising due to 
the singularity of the one-dimensional Coulomb problem can be 
avoided with the use of the generalized Coulomb potential. 
\end{abstract}



\section{Introduction}

Local potentials have been used to model the interactions 
of the subatomic world ever since the introduction of 
quantum mechanics. Some of these (like the Coulomb potential) 
do not differ essentially from the forces observed in nature, 
while most of them (like the harmonic oscillator, for example) 
represent approximations of the actual physical situation. 
The potential shape, defined by the potential type and the 
parameters in it is usually chosen in a way that reflects 
the physical picture our intuition associates with the problem; 
therefore we can define attractive or repulsive, short-range 
or long-range potentials, etc. 

Some of the potentials used in quantum mechanics are 
exactly solvable. This means that the energy eigenvalues, the 
bound-state wave functions and the scattering matrix can be 
determined in closed analytical form. The range of these potentials 
has been extended considerably in the recent years by investigations 
inspired by supersymmetric quantum mechanics (SUSYQM), for example. 
(See Refs. \cite{susyqm} for recent reviews.) 
Most of the solvable potentials are one-dimensional problems, which 
means that they are defined on the $(-\infty,\infty)$ domain (or 
on some finite interval of it), or they can be reduced to radial 
problems in higher spatial dimensions, in which case they are 
confined to the positive real axis. 
The simplest exactly solvable potentials are those belonging to the 
so-called shape-invariant class \cite{gen83}. Many of the most 
well-known potentials, like the Coulomb, harmonic oscillator,  
Morse, P\"oschl--Teller, etc. potentials belong to this class. 
Usually 12 potentials are mentioned to have the property of 
shape-invariance \cite{le89}, although two of them are just 
alternative forms of other shape-invariant potentials. 

A rather more general potential type is that of the Natanzon 
potentials \cite{nat70}, the solutions of which can 
be reduced to a single hypergeometric or confluent hypergeometric 
function. These potentials are written as complicated six-parameter 
expressions of the coordinate, and their energy eigenvalues are 
given by an implicit formula. Although these problems are solvable 
exactly in principle, only some specific Natanzon potentials have been 
studied in detail \cite{gi84,gi85,stb,lgbw93}, because the 
formalism is getting more and more complicated with the increasing 
number of parameters handled. Some of the Natanzon potentials 
contain certain  shape-invariant potentials as special cases: 
the Ginocchio potential \cite{gi85} can be considered as a 
generalization of the P\"oschl--Teller potential, while the 
generalized Coulomb potential \cite{lgbw93} is a simultaneous extension 
of the Coulomb and the harmonic oscillator potentials.     
There are also further potentials beyond the Natanzon class. 
These include, for example quasi-exactly solvable (QES) potentials 
\cite{qes}, conditionally exactly solvable (CES) potential \cite{ces}
and the SUSYQM partners of Natanzon-class potentials \cite{co87}, of 
which certain CES potentials form a subclass \cite{junker}.

The study of exactly solvable potentials represents an interesting 
field of mathematical physics in itself, but results from this area 
are also essential for the description of realistic physical 
problems. More and more interactions can be modeled by making 
advantage of the rather flexible potential shapes offered by 
exactly solvable potentials. The solutions of these can be 
applied directly, or they can be combined with numerical 
calculations. In the simplest case analytical calculations can 
aid numerical studies in areas where numerical techniques might 
not be safely controlled. This is the case, for example, when 
bound-state wave functions with arbitrary node numbers are 
required, for certain singular potentials, or for complex 
potentials \cite{ba96}. As the next level of complexity, 
analytical solutions can supply a basis for numerical calculations. 

Virtually all quantum mechanical methods rely in some respect on
analytically solvable potentials. Very often their wave function 
solutions are used as a Hilbert-space basis.
More powerful methods can be constructed if we select a basis
which allows the exact analytical calculation of the Green's
operator of an analytically solvable potential.
Some time ago by utilizing the Coulomb-Sturmian functions
as basis functions the Coulomb Green's operator was calculated
analytically and a quantum mechanical approximation method was developed
\cite{p-pse}. Recently this method was extended for approaching the
really challenging three-body Coulomb problem \cite{p-fagyi}.
We note here that in this work the exact analytical representation
of the Coulomb Green's operator on the Coulomb--Sturmian basis
was essential. In our recent work \cite{klp97} we showed that
if on some basis representation the Hamiltonian possesses
an infinite tridiagonal matrix (J-matrix)  form then its resolvent
can be given in terms of continued fractions which can be
calculated analytically on the whole complex plane. This theorem was
exemplified by the Coulomb and harmonic oscillator Green's operators.

The aim of the present work is to demonstrate that various techniques 
developed for the simple shape-invariant potentials, like the Coulomb
and harmonic oscillator potentials, can be 
generalized for certain Natanzon-class potentials too without 
making the formalism significantly more involved, and that
these more general potentials can help our understanding of 
both practical and theoretical problems. 
The actual example we consider here is the generalized Coulomb potential 
\cite{lgbw93}, which is the member of the Natanzon confluent 
potential class \cite{cs91}. 
This potential is Coulomb-like asymptotically, while its short-range 
behavior depends on the parameters: it can be finite or singular 
as well at the origin. Its shape therefore can approximate various 
realistic problems, such as nuclear potentials with relatively 
flat central part, or atomic potentials that incorporate the effect of 
inner closed shells by a phenomenological repulsive core. 

An advantageous feature of the generalized Coulomb problem 
is that the equivalent of the Coulomb--Sturmian basis can be defined 
for it, together with all its implications. For example, 
the method of calculating the matrix elements of the Green's 
operator as described in Ref. \cite{klp97} for the Coulomb and
the harmonic oscillator potentials can be extended to this
case too, so that the above two problems appear in
the present treatment as special cases.  

Using a particular differential realization, an SU(1,1) algebra can 
also be associated with this potential, which reduces to known 
SU(1,1) algebras in the Coulomb and harmonic oscillator limits. 

Another important feature of this potential is that it can be 
formulated as a one-dimensional problem, but it can avoid the 
pathologic features of the one-dimensional Coulomb problem, which 
arise due to the $\vert x \vert^{-1}$-like singularity. 
Since this potential can be chosen to be finite at $x=0$, it offers a 
convenient basis for calculations which require the use of a 
one-dimensional Coulomb-like problem. 

The arrangement of this paper is the following. In Section \ref{gcoul} 
we give the basic formalism for the generalized Coulomb problem 
by recapitulating and extending former results. 
Section \ref{res} contains the main results of the paper: 
{\it i)} the discussion of the Coulomb--oscillator connection 
using the two limiting cases of the potential,
{\it ii)} the determination of the matrix elements of the Green's 
operator, {\it iii)} the description of an SU(1,1) algebra 
associated with the problem and {\it iv)} discussion on the 
one-dimensional case, which elucidates the complications related 
to the singularity of the one-dimensional Coulomb problem. 
Finally, we summarize the results and discuss further possibilities 
in Sect. \ref{sum}.

\section{The generalized Coulomb potential in $D$ dimensions}
\label{gcoul} 

Let us consider the Schr\"odinger equation in $D$ spatial 
dimensions with a potential that depends only on the radial 
variable $r$. Separating the angular variables from the 
wave function we obtain the radial Schr\"odinger equation 
\begin{equation}
\hat{H}_0\psi(r)\equiv
\left(
-\frac{ {\rm d}^2 }{ {\rm d} r^2} 
+\frac{1}{  r^2}(l+\frac{D-3}{ 2})(l+\frac{D-1}{ 2}) 
+ v(r) \right) \psi(r)=\epsilon\psi(r)\ , 
\label{sch}
\end{equation}
where $v(r)\equiv 2m\hbar^{-2} V(r)$ and $\epsilon \equiv  
2m\hbar^{-2} E$. The centrifugal term, which depends on the 
angular momentum $l$ originates from the kinetic term, i.e. 
from the $D$-dimensional Laplace operator after separating 
the angular variables. The bound-state wave functions solving
(\ref{sch}) are normalized as
\begin{equation}
\int^{\infty}_0 \vert \psi(r)\vert^2 {\rm d}r =1 \ . 
\label{norm}
\end{equation}

Here we recapitulate the results obtained for the generalized 
Coulomb potential in Ref. \cite{lgbw93} by adapting it to the 
$D$-dimensional case. Changing slightly the notation used 
in Ref. \cite{lgbw93} we define the generalized Coulomb potential 
as   
\begin{eqnarray}
v(r) & = & -\frac{1}{  r^2}(l+\frac{D-3}{ 2})(l+\frac{D-1}{ 2}) 
+ (\beta-\frac{1}{2})(\beta-\frac{3}{2}) \frac{C}{4h(r)(h(r)+\theta)} 
\nonumber \\
& & -\frac{q}{h(r)+\theta} - \frac{3C}{16 (h(r)+\theta)^2} 
+\frac{5C\theta}{16(h(r)+\theta)^3} \ ,  
\label{pot}
\end{eqnarray}
where $h(r)$ is defined \cite{lgbw93} in terms 
of its inverse function 
\begin{equation}
r=r(h)=C^{-\frac{1}{2}}\left( \theta \tanh^{-1}\left(\left(
\frac{h}{h+\theta}\right)^{\frac{1}{2}}\right)
+(h(h+\theta))^{\frac{1}{2}}\right) \ .
\label{rh}
\end{equation}
The $h(r)$ function maps the $[0,\infty)$ half axis onto 
itself and can be approximated with $h(r)\simeq C^{\frac{1}{2}}r$ 
and $h(r)\simeq Cr^2/(4\theta)$ in the $r\rightarrow\infty$ and 
$r\rightarrow 0$ limits, respectively. 
 
Bound states are located \cite{lgbw93} at 
\begin{equation}
\epsilon_n = -\frac{C}{4}\rho_n^2 \ ,
\label{en}
\end{equation}
where
\begin{equation}
\rho_n= 
\frac{2}{\theta}\left(\left((n+\beta/2)^2 +\frac{q\theta}{C}
\right)^{\frac{1}{2}}-(n+\beta/2)\right) \ .
\label{rho}
\end{equation}
The bound-state wave functions can be written in terms of associated 
Laguerre polynomials as 
\begin{equation}
\psi_{n}(r)=C^{\frac{1}{4}}\rho_n^{\frac{\beta+1}{2}}
\left(\frac{\Gamma(n+1)}{ 
\Gamma(n+\beta)(2n+\beta+\rho_n \theta)}\right)^{1/2} 
(h(r)+\theta)^{\frac{1}{4}} (h(r))^{\frac{2\beta-1}{4}}
\exp(-\frac{\rho_n}{2}h(r)) L_n^{(\beta-1)}(\rho_n h(r)) \ .
\label{bswf}
\end{equation}

Potential (\ref{pot}) clearly carries angular momentum dependence: 
its first term merely compensates the centrifugal term arising 
from the kinetic term of the Hamiltonian. Its second term also 
has $r^{-2}$-like singularity (due to $h^{-1}(r)$), and as it 
will be demonstrated later, 
it cancels the angular momentum dependent term in the two 
important limiting cases that recover the $D$-dimensional Coulomb 
and the harmonic oscillator potentials. The third term of (\ref{pot}) 
represents an asymptotically Coulomb-like interaction, while the 
remaining two terms behave like $r^{-2}$ and $r^{-3}$ for large 
values of $r$. 

In the remaining part of this Section we consider the specific 
case of $D=3$ and $l=0$. 
The $S$ matrix of the generalized Coulomb potential can be derived 
in complete analogy with that of the Coulomb problem for $D=3$. 
Although this can only be done exactly for $l=0$, the singular 
term imitating the centrifugal term in Eq. (\ref{pot}) can be 
defined to be part of the potential. Using the 
notation of Ref. \cite{newton} the equivalent of the general 
regular and irregular solutions are 
\begin{equation}
\varphi^{(G)}(k,r)=C^{-\frac{\beta}{4}}
(h(r)+\theta)^{\frac{1}{4}} (h(r))^{\frac{2\beta-1}{4}}
\exp(-\frac{\rho}{2}h(r)) \Phi(\frac{\beta}{2}+{\rm i}\nu ,
\beta ; \rho h(r)) 
\label{regsol}
\end{equation}
and
\begin{equation}
f^{(G)}(k,r)=\rho^{-\frac{\beta}{4}}{\rm e}^{\frac{\nu\pi}{2}}
(h(r)+\theta)^{\frac{1}{4}} (h(r))^{\frac{2\beta-1}{4}}
\exp(-\frac{\rho}{2}h(r)) \Psi(\frac{\beta}{2}+{\rm i}\nu ,
\beta ; \rho h(r)) 
\label{irregsol}
\end{equation}
where $\Phi(a,c;z)$ and $\Psi(a,c;z)$ are the regular and the 
irregular confluent hypergeometric functions \cite{erd}, and 
$\rho$ and $\nu$ have to be chosen as 
\begin{equation}
\rho\equiv \rho(k)=-\frac{2{\rm i}k}{C^{1/2}}, \hspace{0.5cm}
{\rm i}\nu\equiv {\rm i}\nu(k)=\frac{\rho\theta}{4}-\frac{q}{C\rho}\ .
\label{rhonu}
\end{equation}
The functions (\ref{regsol}) and (\ref{irregsol}) reduce to 
$\varphi^{(c)}(k,r)$ and $f^{(c)}(k,r)$ of Ref. \cite{newton} 
in the $\theta\rightarrow 0$ Coulomb limit. The $S$ matrix for 
$l=0$ is then expressed as
\begin{equation}
S_0(k)=(-1)^{\frac{\beta}{2}+1}
\frac{\Gamma(\frac{\beta}{2}+{\rm i}\nu)} 
{\Gamma(\frac{\beta}{2}-{\rm i}\nu)} \ .
\label{smat}
\end{equation}
The extra phase factor appears because of the $r^{-2}$-type 
singular term, which is now defined to be part of the 
potential. In the Coulomb limit this expression becomes part of 
the centrifugal term, which is dealt with separately.

The long-range behavior of potential (\ref{pot}) suggests its use 
in problems associated with the electrostatic field of some charge 
distribution. The deviation from the Coulomb potential close to 
the origin can be viewed as replacing the point-like charge with 
an extended charged object. The relevant charge density is readily 
obtained from the potential using 
\begin{equation}
\rho(r) = -\frac{\hbar^2}{8\pi m{\rm e}} \Delta v(r)\ .
\label{lapl}
\end{equation}
Choosing $D=3$ and neglecting the terms singular at the origin 
by making the choice $l=0$ and $\beta=$1/2 or 3/2 we get 
\begin{eqnarray}
\rho(r) & = & -\frac{\hbar^2}{8\pi m{\rm e}} [
-\frac{2qC}{(h(r)+\theta)^3} 
+\frac{20qC\theta-9C^2}{8(h(r)+\theta)^4} 
+\frac{81C^2\theta}{16(h(r)+\theta)^5} 
-\frac{135C^2\theta^2}{32(h(r)+\theta)^6} 
\nonumber \\
&&+\frac{2C^{1/2}}{r}\left(\frac{h(r)}{h(r)+\theta}\right)^{1/2}
\left( 
\frac{q}{(h(r)+\theta)^2} 
+\frac{3C}{8 (h(r)+\theta)^3} 
-\frac{15C\theta}{16(h(r)+\theta)^4}
\right) ]\ .
\label{chd}
\end{eqnarray}
In Figs. 1 and 2 we present examples for the actual shape of potential 
(\ref{pot}) and the corresponding charge distribution (\ref{chd}) 
for various values of the parameters. It can be seen that 
this potential is suitable for describing the Coulomb field 
of extended objects. It is a 
general feature of potential (\ref{pot}) that for small values of 
$\theta$ a (finite) positive peak appears near the origin, which 
also manifests itself in a repulsive ``hard core'', corresponding 
to a region with positive charge density (see Fig. 2). 
%
%

\section{Discussion}
\label{res}

\subsection{The Coulomb and harmonic oscillator limits}
\label{limits}

As described in Ref. \cite{lgbw93}, the introduction of the 
generalized Coulomb problem was inspired by a simple method 
of transforming the Schr\"odinger equation into the differential 
equation of the confluent hypergeometric function. The 
bound-state solutions of these problems contain associated 
Laguerre polynomials. The three most well-known examples of 
this type are the Coulomb, the harmonic oscillator and the 
Morse potentials, which are all members of the shape-invariant 
potential class \cite{gen83}. These three potentials are 
recovered by specific choices of the variable transformation 
reducing the Schr\"odinger equation to the differential 
equation of the associated Laguerre polynomials. This is also 
reflected by the structure of the function $h(r)$ in Eq. 
(\ref{rh}): when $h(r)$ is proportional to $r$ and $r^2$, one 
obtains the Coulomb problem and the harmonic oscillator 
potential, respectively. These limits can readily be 
realized by specific choices of the parameters in Eq. 
(\ref{rh}): the first one follows from the $\theta\rightarrow 0$ 
limit \cite{lgbw93}, while the second one is reached by taking  
$\theta\rightarrow \infty$, while keeping $C/\theta\equiv \tilde{C}$ 
constant \cite{hoii}. 

Besides taking the $\theta\rightarrow 0$ limit, the Coulomb problem 
in $D$-dimensions is recovered from Eq. (\ref{pot}) by the  
$\beta=2l+D-1$ and $C^{-\frac{1}{2}}q=2mZ{\rm e}^2/\hbar^2$, 
choices: the third term of (\ref{pot}) becomes the Coulomb term, 
the fifth one vanishes, while the other three all become 
proportional with $r^{-2}$ and cancel out completely.   

In order to reach the oscillator limit one also has to 
redefine the potential (\ref{pot}) and the energy eigenvalues 
by adding $q/\theta$ to both. (This is equivalent to 
$\tilde{C} \tilde{D} \theta$ 
in the notation of Ref. \cite{hoii}.) This choice simply represents 
resetting the energy scale: $\epsilon=0$ corresponds to 
$v(r\rightarrow\infty)$ 
for the Coulomb problem, and to $v(r=0)$ for the harmonic oscillator. 
(Note that the energy eigenvalues also have different signs in the two 
cases.) Besides $C/\theta =\tilde{C}$, the $\tilde{q} \equiv q/\theta^2$ 
parameter also has to remain constant in the    
$\theta\rightarrow\infty$ transition here. The potential thus 
adapted to the harmonic oscillator limit reads 
\begin{eqnarray}
\tilde{v}(r) \equiv v(r)+q\theta & = &  
-\frac{1}{  r^2}\left(l+\frac{D-3}{ 2}\right)\left(l+\frac{D-1}{ 2}\right) 
+ \left(\beta-\frac{1}{2}\right)\left(\beta-\frac{3}{2}\right)
\frac{\tilde{C}}{4h(r)(1+\frac{h(r)}{\theta})} 
\nonumber \\
& & -\frac{\tilde{q} h(r)}{1+\frac{h(r)}{\theta}} 
- \frac{3\tilde{C}}{16 \theta} 
\frac{1}{\left(1+\frac{h(r)}{\theta}\right)^2} 
+\frac{5\tilde{C}}{16\theta}\frac{1}{\left(1+\frac{h(r)}{\theta}\right)^3} \ .
\label{potho}
\end{eqnarray}
The harmonic oscillator potential is recovered from (\ref{potho}) 
by the $\beta=l+D/2$ and $\tilde{C} \tilde{q}=(2m\omega/\hbar)^2$ choice. The 
two last terms in (\ref{potho}) vanish, the first and the second 
cancel out, while the third one reproduces the harmonic oscillator 
potential. The new form of the energy eigenvalues is 
\begin{equation}
\tilde{\epsilon}_n \equiv \epsilon_n +q/\theta = 
\tilde{C}(2n+\beta)\left(\left( \frac{1}{\theta^2}(n+\frac{\beta}{2})^2 
+\frac{\tilde{q}}{\tilde{C}}\right)^{\frac{1}{2}} -\frac{1}{\theta}
(n+\frac{\beta}{2})\right) \ ,
\label{enho}
\end{equation}
which indeed, reduces to the $\tilde{\epsilon}_n=(2m\omega/\hbar)(2n+l+D/2)$ 
oscillator spectrum in the $\theta\rightarrow\infty$ limit. 
The wave functions (\ref{bswf}) are unchanged, except for the 
redefinition of the parameters. 

Note that the generalized Coulomb problem establishes a link 
between the Coulomb problem and the harmonic oscillator 
potential in different spatial dimensions. This is best seen 
by inspecting the wave functions (\ref{bswf}). If the $\beta-1$ 
parameter of 
the associated Laguerre polynomial is required to be the same 
in the two limits, we get an interrelation between the value 
of the angular momentum and the spatial dimension to be used 
for the Coulomb and the harmonic oscillator case:
\begin{equation}
l^O + \frac{D^O}{2} =2l^C + D^C -1 \ .
\label{ld}
\end{equation}
Considering $l^O=2l^C$, Eq. (\ref{ld}) implies $D^O=2D^C-2$ which
establishes link between the $(D^C,D^O)=$(2,2), (3,4),
(4,6), (5,8), $\dots$ etc. pairs. (Of these the (3,4) pair
corresponds to the Kustaanheimo--Stiefel transformation
\cite{ks}.) This case is called the ``direct map'' between 
the Coulombic and oscillator solutions in Ref. \cite{knt85}, 
while with the $l^O=2l^C+\lambda$ choice the ``general map'' 
can be recovered. 
These results suggest that 
the generalized Coulomb potential can be used to formulate  
a continuous transition between the Coulomb and harmonic 
oscillator potentials, as opposed to the usual procedure 
that employs a unique variable and parameter transformation 
to reach this goal.

\subsection{The generalized Coulomb--Sturmian basis and the 
matrix elements of the Green's operator}
\label{green}

The generalized Coulomb--Sturmian equation which depends on 
$n$ as a parameter, has similar
structure  
to the eigenvalue equation in Eq. (\ref{sch}) with potential 
(\ref{pot})
\begin{eqnarray}
\hat{X}\phi(\rho,r) & \equiv &
[
-\frac{{\rm d}^2}{{\rm d} r^2} 
-\frac{3C}{16 (h(r)+\theta)^2} 
+\frac{5C\theta}{16(h(r)+\theta)^3} 
+\frac{C(\beta-\frac{1}{2})(\beta-\frac{3}{2})}{4h(r)(h(r)+\theta)} 
\nonumber \\
&&-\left(\frac{\rho^2\theta}{4}+\rho(n+\frac{\beta}{2})\right)
\frac{C}{h(r)+\theta} + \frac{C}{4}\rho^2 ] \phi(\rho,r)=0\ ,
\label{gcst}
\end{eqnarray}
and is solved by the generalized Coulomb--Sturmian (GCS) functions 
\begin{equation}
\langle r \vert n \rangle \equiv 
\phi_{n}(\rho,r)  =\left( \frac{\Gamma(n+1)}{ 
\Gamma(n+\beta)}\right)^{1/2} (\rho h(r)+\rho\theta)^{\frac{1}{4}} 
(\rho h(r))^{\frac{2\beta-1}{4}} \exp(-\frac{\rho}{2}h(r))
L_n^{(\beta-1)}(\rho h(r)) \ . 
\label{gcsf}
\end{equation}
Here $\rho$ is a parameter 
characterizing the generalized Coulomb--Sturmian basis. The GCS 
functions, being solutions of a Sturm-Liouville problem,
 have the property of being orthonormal with 
respect to the weight function $C^{\frac{1}{2}}(h(r)+\theta)^{-1}$. 
Introducing the notation $\langle r \vert \widetilde n \rangle 
\equiv \phi_n(\rho,r) C^{\frac{1}{2}}(h(r)+\theta)^{-1}$ 
the orthogonality and completeness relation of the GCS functions 
can be expressed as 
\begin{equation}
\langle n \vert \widetilde n \rangle =\delta_{n n'} 
\label{ort}
\end{equation}
and 
\begin{equation}
1=\sum_{n=0}^{\infty} \vert \widetilde n \rangle \langle n \vert 
=\sum_{n=0}^{\infty} \vert  n \rangle \langle \widetilde n \vert \ . 
\label{comp}
\end{equation}

Straightforward calculation shows that both the overlap of 
two GCS functions and the $\langle n' \vert \hat{H}_0 \vert n \rangle $ 
matrix element can be expressed as a tridiagonal matrix, 
therefore the matrix elements of the $\epsilon-\hat{H}_0$ operator 
also have this feature:  
\begin{eqnarray}
\langle n \vert \epsilon-\hat{H}_0 \vert n' \rangle & = & 
\delta_{nn'}\left( \frac{\epsilon}{C^{\frac{1}{2}}\rho} (2n+\beta -
\rho\theta ) -\frac{C^{\frac{1}{2}}\rho}{4} \left( -\frac{4q}{C\rho} 
+(2n+\beta)  \right) \right)
\nonumber \\
& - & \delta_{n n'+1} \left( n(n+\beta-1)\right)^{\frac{1}{2}} 
\left(\frac{\epsilon}{C^{\frac{1}{2}}\rho} + \frac{C^{\frac{1}{2}}
\rho}{4}\right) 
-\delta_{n n'-1} \left( (n+1)(n+\beta)\right)^{\frac{1}{2}} 
\left(\frac{\epsilon}{C^{\frac{1}{2}}\rho} + \frac{C^{\frac{1}{2}}
\rho}{4}\right) \ . 
\label{eminh0}
\end{eqnarray}
This means that similarly to the $D$-dimensional Coulomb and 
harmonic oscillator potential the matrix elements of the 
Green's operator can be determined by using continued fractions, 
as described in Ref. \cite{klp97}. 
The present results, therefore, extend the applicability of this method 
to a new potential problem. The formulae presented here reduce to 
those in \cite{klp97} in the appropriate limits discussed in 
Subsection {limits}. The 
role of the Coulomb-Sturm parameter $b$ used in \cite{klp97} is 
now played by $C^{\frac{1}{2}} \rho /2$.

\subsection{Algebraic aspects}
\label{alg}

Here we show that an SU(1,1) algebra can be associated with
the generalized Coulomb problem. The generators of this algebra are 
\begin{eqnarray}
&&\hat{J}_3=\frac{h+\theta}{C\rho}\hat{X}+(n+\frac{\beta}{2})\ ,
\hspace{0.5 cm}
\hat{J}_1=\hat{J}_3-\frac{\rho}{2}h \ ,
\nonumber \\
&& \hat{J}_2=
-\frac{\rm i}{C^{1/2}}(h(h+\theta))^{1/2}\frac{{\rm d}}{{\rm d} r} 
-\frac{{\rm i}\theta}{4(h+\theta)}\ ,
\label{su11}
\end{eqnarray}
an they satisfy the commutation relations 
\begin{equation}
[\hat{J}_1,\hat{J}_2]=-{\rm i}\hat{J}_3 \hspace{0.5cm} 
[\hat{J}_2,\hat{J}_3]={\rm i}\hat{J}_1 \hspace{0.5cm} 
[\hat{J}_3,\hat{J}_1]={\rm i}\hat{J}_2\ . 
\label{su12}
\end{equation}
As can be seen from Eq. (\ref{gcst}), $\hat{J}_3$ is diagonal in the 
basis (\ref{gcsf}) with eigenvalues $m=n+\frac{\beta}{2}$.  
The elements of this basis can then be associated with the 
infinite-dimensional discrete unitary irreducible representation 
of SU(1,1) called \cite{wyb} discrete principal series $D^+_j$, for 
which the allowed values of $m$ are 
\begin{equation}
m=-j, -j+1, -j+2, \dots
\label{m}
\end{equation}
with $j$ being negative. It is natural then to identify $j$ as 
$j=-\frac{\beta}{2}$. Direct calculations show that the ladder 
operators connect the neighboring members of this basis: 
\begin{equation}
\hat{J}_+ \phi_n(\rho,r) \equiv 
(\hat{J}_1 + {\rm i}\hat{J}_2) \phi_n(\rho,r) =
[(n+1)(n+\beta)]^{1/2} \phi_{n+1}(\rho,r)\ ,
\label{jp}
\end{equation}
\begin{equation}
\hat{J}_- \phi_n(\rho,r) \equiv 
(\hat{J}_1 - {\rm i}\hat{J}_2) \phi_n(\rho,r) =
[n(n+\beta-1)]^{1/2} \phi_{n-1}(\rho,r)\ .
\label{jm}
\end{equation}

We find that the eigenvalues of the Casimir invariant 
\begin{equation}
\hat{C}_2=\hat{J}_3^2-\hat{J}_1^2-\hat{J}_2^2
\label{c2}
\end{equation}
are $\frac{\beta}{2}(\frac{\beta}{2}-1)=j(j+1)$, as expected, 
and that they set the strength of the fourth term in (\ref{gcst}). 
For $\theta\neq 0$ this is the only singular term and it 
behaves like $\gamma r^{-2}$ with 
$\gamma=(\beta-\frac{1}{2})(\beta-\frac{3}{2})=4j(j+1)+\frac{3}{4}$. 
It is interesting to inspect the allowed values of $\gamma$ for the 
different unitary irreducible representations of SU(1,1). For the 
discrete principal series $D^+_j$ $\beta > 1$ holds, which always 
secures $-\frac{1}{4} < \gamma $, i.e. the potential has 
repulsive or ``weakly attractive'' \cite{sing} $r^{-2}$-type 
singularity. For the supplementary series \cite{wyb} 
$-\frac{1}{2} < j < 0$ holds, which results in 
$-\frac{1}{4} < \gamma < \frac{3}{4}$. This is exactly the domain 
where both independent solutions are square integrable at the 
origin \cite{sing}: for $0 < \gamma < \frac{3}{4}$ one of these 
vanishes at $r=0$ and the other one is infinite there, while for 
$-\frac{1}{4} < \gamma < 0$ for both solutions vanish at $r=0$. 
>From (\ref{gcsf}) it is seen that solutions regular and 
irregular at the origin correspond to $\beta > \frac{1}{2}$ and 
$\beta < \frac{1}{2}$. This seems to indicate that one regular 
solution (with $\beta > 1$) is associated with $D^+_j$, while the 
second square integrable solution, which is either regular or 
infinite at the origin (depending on $\beta$) might be related to 
the supplementary series, for which $0 < \beta < 1$ holds. For the 
sake of completeness we note that for the continuous series 
\cite{wyb} $C^0_k$ and $C^{1/2}_k$, $j=-\frac{1}{2}+{\rm i}k$ 
($k > 0$, real) is valid, which results in the strongly 
singular $\gamma < -\frac{1}{4}$ case. The solutions then 
oscillate infinitely near the origin and are unbounded from 
below \cite{sing}, which can be interpreted as the falling of 
the particle into the center of attraction \cite{landau}. 
The situation is similar to that described in Ref. \cite{wig95} 
for the $v(x)=\gamma\sin^{-2} x$ potential: the various 
unitary irreducible representations of the SU(1,1) spectrum 
generating group there also corresponded to different types 
of singularities. 

We note that we analyzed the singularities of  
the generalized Coulomb--Sturmian equation (\ref{gcst}) and its 
solutions (\ref{gcsf}), but similar considerations of the physical 
potential (\ref{pot}) and its solutions can also be performed
taking $D=3$ and $l=0$. The algebraic construction, however, 
does not apply to this latter problem. This is because the bound-state 
solutions (\ref{bswf}) pick up extra $n$-dependence through $\rho_n$, 
which is not accessible for the ladder operators otherwise changing 
$n$ as in Eqs. (\ref{jp}) and (\ref{jm}). 

The present realization of the SU(1,1)
algebra is a special case of that described in Ref.~\cite{cs91}
in relation with the Natanzon confluent potentials. 
Considering the Coulomb and harmonic oscillator limits discussed in 
Subsection \ref{limits} and setting the dimension to $D=3$, the 
generators reduce to the forms presented for the two problems 
separately in Ref.~\cite{cooper}. We note that the spectrum 
generating algebra associated this way with the radial 
harmonic oscillator problem in three dimensions is different 
from the two-parameter realization of the SU(1,1) algebra 
discussed in Refs. \cite{le94,goslar}, because the ladder operators 
there are linear differential operators and the Hamiltonian is related 
to the Casimir invariant, while here the ladder operators 
are second-order differential operators and the Hamiltonian is 
essentially a linear function of generator $\hat{J}_3$. 

Finally, we note that similar limiting cases of other Natanzon-class 
potentials can  also be treated in terms of SU(1,1) algebras. 
In Ref. \cite{goslar} an SU(1,1) algebra associated with the 
Ginocchio potential \cite{gi84} was analyzed, and it was shown 
that in two different limiting cases, which both result in the 
P\"oschl--Teller potential, two different algebras can be 
recovered from the original one: the first being a spectrum 
generating algebra and the second a potential algebra.

\subsection{The one-dimensional case}   
\label{1d}

The formalism developed previously is valid for $D=1$ too, 
nevertheless, some particular properties of one-dimensional 
problems justify a separate treatment of this case. 
First, the implicit definition of the $h(r)$ function in 
Eq. (\ref{rh}) has to be extended to negative values of 
$r$, which we now denote with $x$. Using the notation of 
(\ref{rh}), we can write that $x=r(h)$ for $x\ge 0$ and 
$x=-r(h)$ for $x<0$. The normalization of the wave functions 
in (\ref{bswf}) also has to be modified with a factor of 
$2^{-1/2}$, accounting for the fact that the integration 
now runs from $-\infty$ to $\infty$, but since the potential 
is symmetric, the bound-state wave functions are either 
even or odd, therefore $\vert \psi(x) \vert^2$ is an even 
function of $x$.   

For one-dimensional problems it is natural to set $l$ to 
0 besides $D=1$, which eliminates the centrifugal term in 
(\ref{pot}). Furthermore, 
in order to avoid $r^{-2}$-like singularities at $x=0$
the second term in (\ref{pot}) also has to be canceled by 
setting $\beta$ to either $1/2$ or to $3/2$. 
Elementary calculations show that the latter choice 
corresponds to bound-state wave functions that vanish at 
$x=0$, and essentially represent physical solutions 
of the problem in higher dimensions as well, while the 
former choice recovers solutions that do not vanish in 
general at $x=0$. These two possibilities can naturally be 
interpreted as odd and even solutions of the one-dimensional 
potential problem. Furthermore, for $x\geq 0$ the two types of 
wave functions can be rewritten into a common notation (up to a 
sign) by making use of the 
relation of associated Laguerre and Hermite polynomials, 
when the former ones have $\alpha=1/2$ or $\alpha=-1/2$ as 
parameters \cite{as70}: 

\begin{equation}
\psi^{(D=1)}_{N}(x)=\frac{C^{\frac{1}{4}}\tilde{\rho}_N^{\frac{3}{4}}}
{2^N\left(\Gamma(\frac{N+1}{2})\Gamma(\frac{N}{2}+1)
(N+\frac{1}{2}+\theta\tilde{\rho}_N\right)^{\frac{1}{2}}}
(h(x)+\theta)^{\frac{1}{4}} \exp(-\frac{\tilde{\rho}_N}{2}h(x)) 
H_N((\tilde{\rho}_N h(x))^{\frac{1}{2}}) \ .
\label{1dwf}
\end{equation}
For $x\leq 0$the bound-state wave functions satisfy 
\begin{equation}
\psi^{(D=1)}_N(-x)=\left\{
\begin{array}{ll} 
\psi^{(D=1)}_N(x) & \mbox{for $N=2n$ ($\beta=\frac{1}{2}$)} \\
-\psi^{(D=1)}_N(x) & \mbox{for $N=2n+1$ ($\beta=\frac{3}{2}$)} \ . 
\end{array}
\right.
\label{1dnegx}
\end{equation}
In (\ref{1dwf}) we defined $\tilde{\rho}_N$ as 
\begin{equation}
\tilde{\rho}_N\equiv \frac{1}{\theta}\left(\left( (N+\frac{1}{2})^2 
+4\frac{q\theta}{C} \right)^{\frac{1}{2}} -(N+\frac{1}{2})
\right) \ ,
\label{1drho}
\end{equation}
which reduces to $\rho_{\left[ \frac{N}{2} \right]}$, where 
even and odd values of $N$ have to be chosen with $\beta=1/2$ and 
$\beta=3/2$, respectively, and the integer part of $N/2$ 
corresponds to $n$ used in $\rho_n$ in Eq. (\ref{rho}). 

An interesting aspect of this potential is that it 
remains finite at $x=0$ ($v(0)=-q/\theta +C /(8\theta^2)$) 
for any finite value of $\theta$, however, 
a narrow, finite peak appears  in the $\theta\rightarrow 0$ 
limit, which then becomes an attractive $-3/(16r^2)$-like 
singularity in the 
Coulomb limit. This is due to the fourth term in (\ref{pot}) and 
it corresponds to a ``weak'' singularity in the sense that 
the center of attraction is not strong enough for the particle 
to become infinitely bound \cite{landau}. This finite barrier 
arising for small, but finite $\theta$ values also introduces the 
possibility of studying tunneling effects in symmetric potential 
wells. We also note that besides being finite at $x=0$,
potential (\ref{pot}) has continuous derivative there, as can 
directly be verified. 

Based on these features the $D=1$ version of potential (\ref{pot}) 
can be used to analyze the peculiarities of the one-dimensional 
Coulomb potential defined as $v^{(c)}=-{\rm e}^2/\vert x\vert$. 
This problem has been the subject of 
intensive studies in the past couple of decades, but there 
is still some controversy in the interpretation of the results
(see e.g. \cite{gor97} for a recent review).  
The unusual features attributed to this singular problem include 
degenerate eigenvalues \cite{lou59} interpreted in terms 
of a hidden O(2) symmetry \cite{dav87}, an infinitely bound ground 
state \cite{lou59} and continuous bound-state spectrum \cite{hai69}.
The last two of these were later found to be based on unacceptable 
solutions of the Schr\"odinger equation \cite{and66,and76}, while 
the unexpected degeneracy was explained by an impenetrable 
barrier at $x=0$, which separates the problem into two disjoint,  
non-communicating systems with identical energy spectra \cite{and76} 
and makes even the concept of parity obsolete here \cite{ose93}. 
Most authors discussing the one-dimensional Coulomb problem 
agree that the usual techniques of quantum mechanics alone in dealing 
with potentials are not sufficient in this case. In Refs. \cite{fis95} 
for example self-adjoint extension of the relevant differential operator 
has been discussed. 

The one-dimensional version of potential (\ref{pot}) can be 
chosen such that it becomes close to non-singular potentials used 
in the approximation of the true one-dimensional Coulomb potential. 
In fact, with appropriate choice of $q$ and $\theta$ any desired 
Coulomb asymptotics and $v(x=0)$ value can be generated. 
Fig. 1 shows potentials with rounded-off shape near $x=0$ 
($\theta=1$, $q=0.5$) and also ones close to the Coulomb potential 
with a cutoff $-{\rm e}^2/(\vert x\vert + a)$ ($\theta=1$, $q=2.5$). 
In contrast with these modified Coulomb potentials, all calculations 
can be performed exactly with (\ref{pot}). 
In the $\theta\rightarrow 0$ limit the generalized Coulomb 
potential recovers the one-dimensional Coulomb potential 
supplemented with the $-\frac{3}{16}x^{-2}$ term. This means that 
the one-dimensional Coulomb potential cannot be 
reached exactly, nevertheless, reasonable approximations of 
it can be given, as it will be demonstrated below.

The odd solutions, of course, vanish 
at $x=0$, while the even solutions have non-zero value there 
as long as $\theta >0$ holds. In the $\theta\rightarrow 0$ limit
$\psi^{(D=1)}_{N=2n}(0)$ varies with $\theta^{1/4}$, so the 
even solutions also tend to zero at $x=0$. This is in accordance 
with the discussion in Subsection \ref{alg} on weakly attractive 
$\gamma r^{-2}$ type singular potentials on the half line: for 
$-\frac{1}{4} < \gamma < 0$ both independent solutions vanish at the 
origin, so the wave functions are necessarily zero at $r=0$. 
If we try to extend the $N=2n$ solutions (\ref{1dwf}) in the 
$\theta=0$ Coulomb limit to the $(-\infty,0)$ domain we find 
that due to its $x^{1/4}$ type behavior at the origin the  
derivative of an even wave function would not be continuous 
anymore. 

It is also interesting to compare the Coulomb limit of the 
one-dimensional generalized Coulomb potential with the true 
one-dimensional Coulomb potential. Switching off gradually the 
$\gamma r^{-2}$ term we find that the two types of solutions 
contain confluent hypergeometric functions of the type $\Phi(-n,c;z)$ 
with $c=1+(1+4\gamma)^{1/2}$ and $c=1-(1+4\gamma)^{1/2}$. Starting 
with $\gamma=-\frac{3}{16}$ we recover the $c=\frac{3}{2}$ and 
$\frac{1}{2}$ corresponding to the two $c=\beta$ values assigned to the 
allowed solutions of the one-dimensional generalized Coulomb potential 
and its Coulomb limit. Taking the $\gamma\rightarrow 0$ limit
results in $c=2$ and $c=0$. The first one of these corresponds to 
the allowed solutions of the one-dimensional Coulomb potential 
(see e.g. Ref. \cite{new94}), while the latter one has to be 
rejected, because the confluent hypergeometric function 
$\Phi(a,c;z)$ is not defined for $c=0$. This shows that even 
solutions which are allowed for the generalized Coulomb 
potential in one dimension do not have equivalents in the 
true one-dimensional Coulomb case. 

We also note that the true one-dimensional Coulomb potential 
can be reached in the $\theta\rightarrow 0$ limit setting 
$\beta$ to 2 from the beginning. In this case, however, a repulsive 
$3/(16r^2)$-type singularity appears, which cancels out only for 
$\theta=0$. This repulsive singularity then separates the problem 
into two parts (with $x > 0$ and $x < 0$) from the beginning, so the 
potential is not a real one-dimensional one. The $\beta=0$ choice, 
in principle, also gives the same potential, however, the solutions 
associated with $\beta=0$ are not defined, as we have shown 
previously. Using $\beta=\varepsilon$ or $2-\varepsilon$ with 
$\varepsilon$ being small we can get as close to the true 
one-dimensional Coulomb potential as we wish, but the repulsive 
singularity is then present for all $\theta$ values. 

The reflection and transmission coefficients can be analyzed 
using the asymptotic behavior of the general solutions of 
the one-dimensional problem. These can be chosen to be even 
and odd functions of $x$. Making use of Eq. (\ref{regsol}) the 
even and the odd solutions can be defined for $x\geq 0$ in  
terms of $\varphi^{(G)}(k,x)$ for $x\geq 0$, setting 
$\beta=\frac{1}{2}$ and $\frac{3}{2}$, respectively; and their 
extension to $x\leq 0$ can be given using a formula 
similar to Eq. (\ref{1dnegx}).
The two solutions are interrelated by Eq. 6.3(3) of Ref. 
\cite{erd}. Due to the symmetric nature of the one-dimensional 
potential ($v(x)=v(-x)$) it is enough to analyze the 
asymptotic behavior of the solutions for $x\rightarrow\infty$: 
the $x\rightarrow -\infty$ case follows naturally. 
Straightforward calculations show that the reflection 
coefficient is
\begin{equation}
R(k)=\frac{{\rm e}^{-{\rm i}\pi/4}}{2}
\left(
\frac{\Gamma(\frac{1}{4}+{\rm i}\nu)} 
{\Gamma(\frac{1}{4}-{\rm i}\nu)} -{\rm i}
\frac{\Gamma(\frac{3}{4}+{\rm i}\nu)} 
{\Gamma(\frac{3}{4}-{\rm i}\nu)} 
\right) \ .
\label{refl}
\end{equation}
Strong reflection is found for potentials having a (finite) 
barrier in $x=0$ (like those in Fig. 1 with $\theta=0.01$), 
while more regular shapes (like that in Fig. 1 with $\theta=1$ 
and $q=2.5$, for example) give weak reflection. The $k$ 
corresponding to maximal reflection increases  as $\theta$ 
get closer to 0 (i.e. as the barrier at $x=0$ gets higher). 
$\vert R(k)\vert$ is small for very large and very small values 
of $k$. Fig. 3 presents illustration for $\vert R(k)\vert^2$ 
for various values of $\theta$. 
Our findings on the reflection coefficient of the one-dimensional 
generalized Coulomb potential seem to support the existence of 
the space splitting effect \cite{ose93} valid for the Coulomb 
potential on one dimension: in situations close to the true 
Coulomb case (e.g. $\theta\rightarrow 0$) $\vert R(k) \vert^2$ 
is close to 1, which corresponds to a nearly impenetrable 
barrier at $x=0$. 
This is also confirmed by our analysis on other approximations 
of the one-dimensional Coulomb potential, i.e. those with 
$\beta=\varepsilon$ or $2-\varepsilon$ above. 

Besides its mathematical aspects the one-dimensional Coulomb 
potential has physical relevance too, in the description of the 
hydrogen atom in strong magnetic field \cite{rud94}, for example. 
In such practical calculations it is reasonable to 
use a non-singular Coulomb-like potential instead of the 
true one-dimensional Coulomb potential: the finite size of 
the nucleus can be a justification for this. This means that 
the one-dimensional Coulomb potential might not be sufficient 
in such calculations: the basis defined with it simply does not 
contain even-parity states. In practical calculations therefore 
the use of bases like that assigned to the generalized Coulomb  
potential is necessary.

\section{Summary and outlook}
\label{sum}

We analyzed the generalized Coulomb potential and extended
results for the $D$-dimensional Coulomb and harmonic oscillator 
potentials to this potential too, which includes these two simple 
problems as special cases. The results presented here have 
both theoretical and practical significance. The theoretical 
aspects concern the treatment of Coulomb-like problems in 
one dimension. Due to the singular nature of the one-dimensional 
Coulomb potential the usual methods of defining a physically 
relevant basis are not sufficient in this case. The generalized 
Coulomb potential avoids these complications and it supplies 
a basis (including both even and odd functions) which can be  
used instead of the Coulomb-Sturmian basis, which is otherwise 
a rather helpful tool in higher dimensions, e.g.\ 
the three-body Coulomb problem has been approached this way
\cite{p-fagyi}. We also gave a 
unification of the SU(1,1) algebras associated with 
the Coulomb and harmonic oscillator problems separately. 

The practical aspects of our work include the introduction of 
the generalized Coulomb-Sturmian basis, which we used 
as a basis for analytic representation of the  Green's operator.  
 The generalized Coulomb potential 
(being the member of the rather general Natanzon potential 
class) offers potential shapes that can approximate various 
physical problems more accurately than the conventional 
textbook potentials. In particular, its long-range Coulomb-like 
behavior makes it a good candidate for describing the 
Coulomb field of extended (non-point-like) charge distributions. 
It can be applied either directly, or as a basis for numerical 
calculations. Its direct applications can be envisaged in 
describing muonic atoms, where it can model the Coulomb field 
of the nucleus. It can also be combined with short-range 
potentials in the potential separable expansion
formalism \cite{p-pse,p-fagyi}, for example. 
Then the Coulombic asymptotics of the 
interaction between the strongly interacting charged objects 
could be taken care of by the Green's operator and
only the short-range (nuclear) interaction would have to be 
approximated. 

Similar extensions of other relatively simple (shape-invariant) 
potentials also seem possible. The example of the generalized Coulomb 
potential shows that the techniques developed for these 
well-known potentials can be extended to more general 
potentials without much technical difficulty. The generalized 
potentials ``inherit'' many important features of the 
simple potentials, of which they are extensions, and, at 
the same time, they can be adapted to more complex physical 
situations, due to their more flexible shape. There are 
other examples confirming the expectations that seemingly 
complicated formulae can be evaluated analytically even for 
highly non-trivial potentials. In Ref. \cite{lbs97}, for example, 
potentials phase-equivalent with the Ginocchio potential 
\cite{gi85} have been determined analytically, using the 
transformations of supersymmetric quantum mechanics. 

Based on the present results we believe that the Natanzon-class 
potentials, which have been analyzed almost exclusively 
in theoretical studies until now, can be used in practical 
applications too. 

\section{Acknowledgements} 
This work has been supported by the OTKA contracts No. 
F20689 and T17298, and grant No. JF 345/93 
of the U.S--Hungarian Science and Technology Joint Fund.


\begin{figure}
\psfig{file=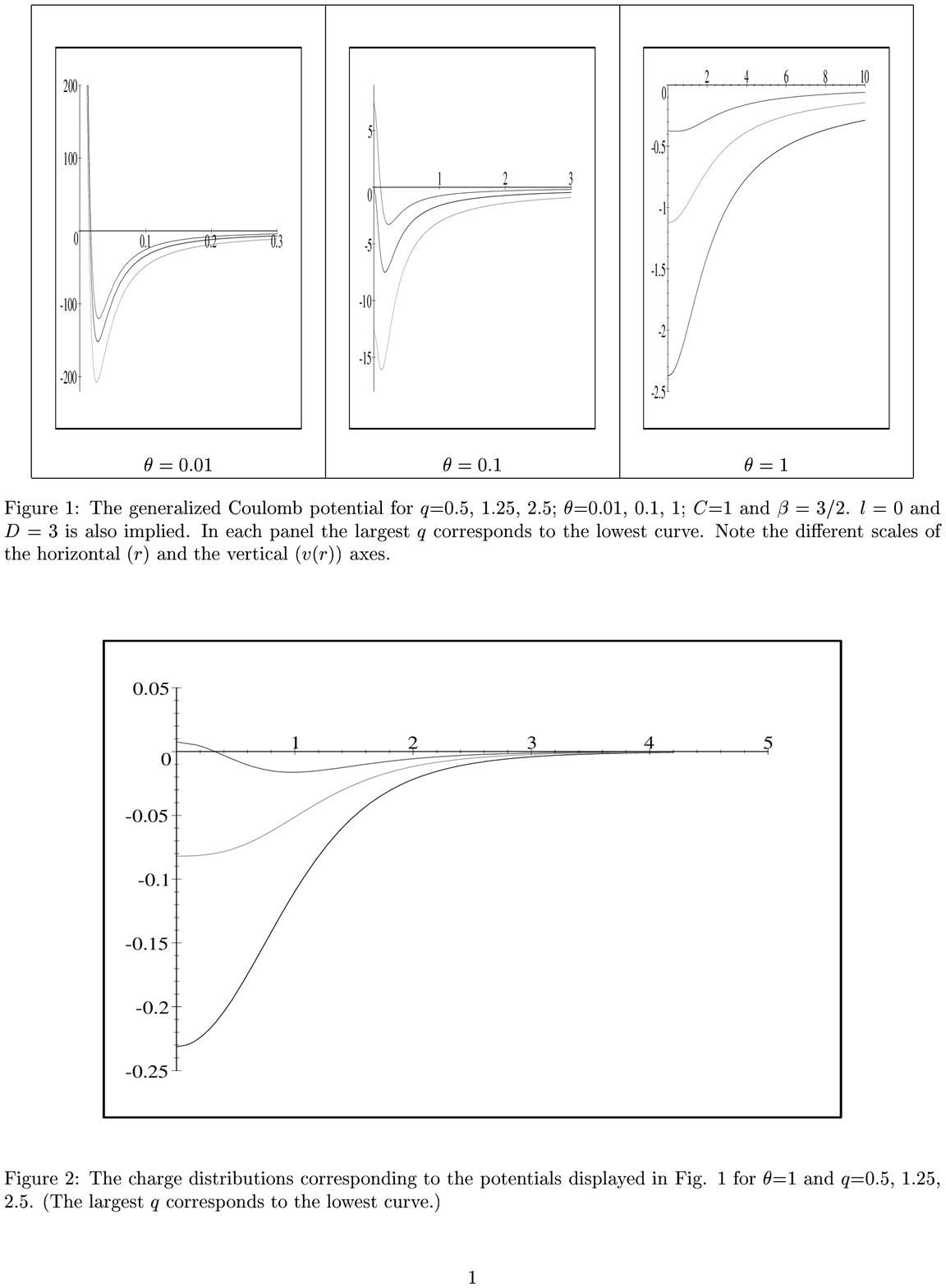,width=15.0cm}
\end{figure} 

\begin{figure}
\psfig{file=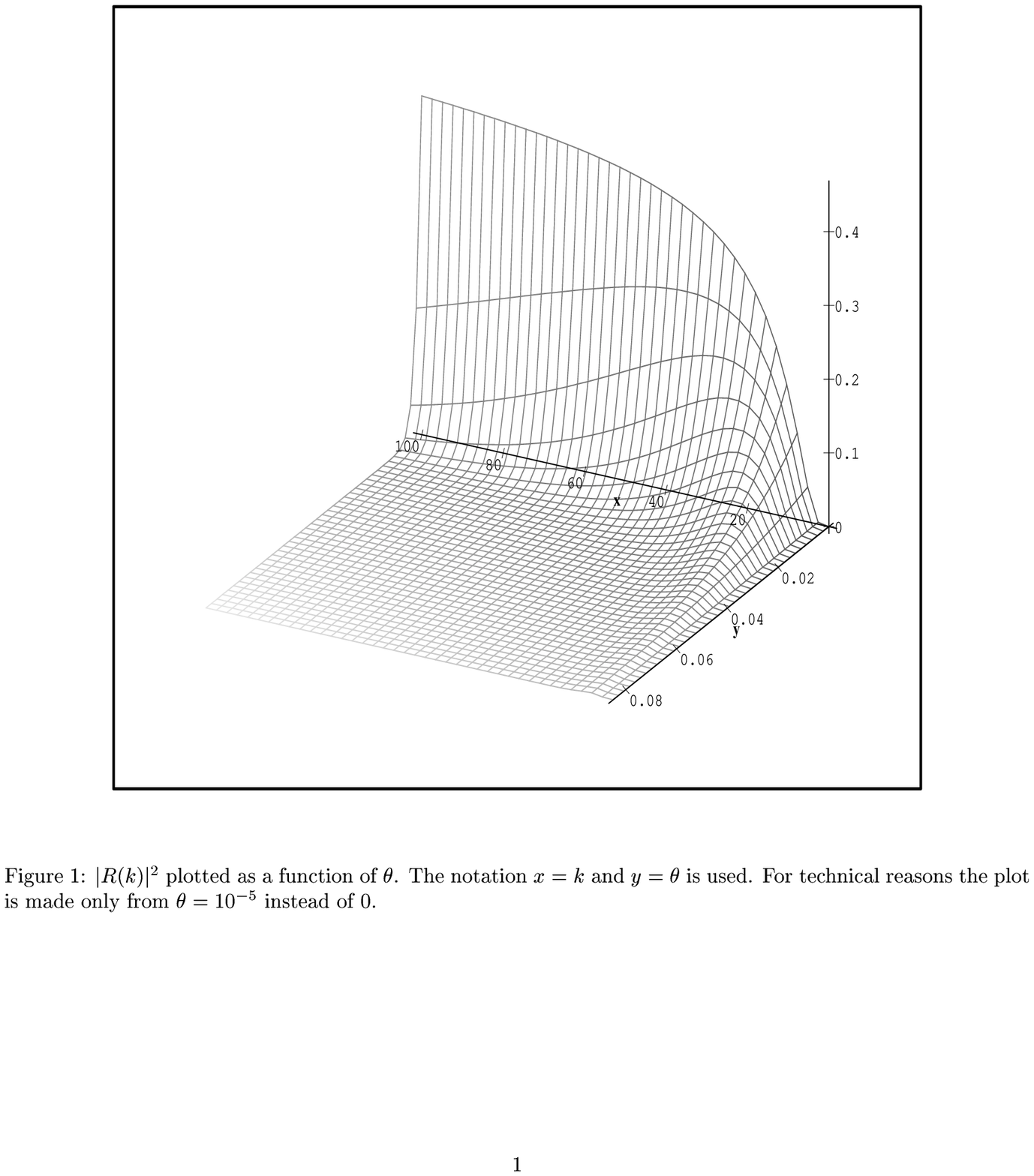,width=15.0cm}
\end{figure}

\end{document}